\documentclass{emulateapj}
\usepackage{amssymb}
\usepackage{amsmath}
\usepackage{mathrsfs}
\usepackage{amsfonts}
\usepackage{textcomp}
\usepackage{color}
\definecolor{mycol}{cmyk}{0.0, 0.11, 0.12, 0.0}
\definecolor{mycol}{cmyk}{0.11, 0.99, 0.11, 0.0}

\def\apjl{\mbox{ApJL}}
\def\apjs{\mbox{ApJS}}

\def\aap{\mbox{A\&A}}

\def\deg{^\circ}

\def\kms{km~s$^{-1}$}

\def\xco{$X_{\rm CO}$ }
\def\xhcn{$X_{\rm HCN}$ }

\def\kmskpc{$\mbox{km s}^{-1}\mbox{ kpc}^{-1}$}

\def\deg{^\circ}

\def\farcs{\hbox{$.\!\!^{\prime\prime}$}}

\newcommand{\Msun}{M$_\odot$}

\newcommand{\COJ}{CO $J(3\to 2)$}

\newcommand{\HCO}{HCO$^+$}
\newcommand{\HCOJ}{HCO$^+$ $J(4\to 3)$}
\newcommand{\HCNRatio}{HCN $J(4\to 3)$ / HCN $J(1\to 0)$}
\newcommand{\HCN}{HCN}
\newcommand{\HCNJ}{HCN $J(4\to 3)$}

\shortauthors{Fathi et al.}
\slugcomment{Accepted for publication in The Astrophysical Journal Letters, June 2, 2015}

\begin{document}

\shorttitle{Local instability signatures in ALMA observations of dense gas in NGC\,7469}
\title{Local instability signatures in ALMA observations of dense gas in NGC\,7469}

\author{Kambiz Fathi$^{1,2}$}
\author{Takuma Izumi$^{3}$} 
\author{Alessandro B. Romeo$^{4}$}
\author{Sergio Mart\'{\i}n$^{5}$} 
\author{Masatoshi Imanishi$^{6,7,8}$}
\author{Evanthia Hatziminaoglou$^{9}$}
\author{Susanne Aalto$^{4}$}
\author{Daniel Espada$^{10,6,7}$} 
\author{Kotaro Kohno$^{3,11}$}
\author{Melanie Krips$^{5}$} 
\author{Satoki Matsushita$^{12}$}
\author{David S. Meier$^{13}$}
\author{Naomasa Nakai$^{14}$}
\author{Yuichi Terashima$^{15}$}
\affil{$^{1}$Stockholm Observatory, Department of Astronomy, Stockholm University, 106 91 Stockholm, Sweden}
\affil{$^{2}$Oskar Klein Centre for Cosmoparticle Physics, Stockholm University, 106 91 Stockholm, Sweden}
\affil{$^{3}$Institute of Astronomy, The University of Tokyo, 2-21-1 Osawa, Mitaka, Tokyo 181-0015, Japan}
\affil{$^{4}$Department of Earth and Space Sciences, Chalmers University of Technology, SE-41296 Gothenburg, Sweden}
\affil{$^{5}$Institut de Radio Astronomie Millim\'etrique, Domaine Univ., 300 Rue de la Piscine, 38406 Saint Martin d'Heres, France}
\affil{$^{6}$National Astronomical Observatory of Japan, 2-21-1 Osawa, Mitaka, Tokyo 181-8588, Japan}
\affil{$^{7}$The Graduate University for Advanced Studies (SOKENDAI), 2-21-1 Osawa, Mitaka, Tokyo 181-8588, Japan}
\affil{$^{8}$Subaru Telescope, NAOJ, 650 North A'ohoku Place, Hilo, HI 96720, USA}
\affil{$^{9}$ESO, Karl-Schwarzschild-Str. 2, 85748 Garching bei M\"unchen, Germany}
\affil{$^{10}$Joint ALMA Observatory, Alonso de Cordova, 3107, Vitacura, Santiago 763-0355, Chile}
\affil{$^{11}$Research Center for the Early Universe, The University of Tokyo, 7-3-1 Hongo, Bunkyo, Tokyo 113-0033, Japan}
\affil{$^{12}$Academia Sinica, Institute of Astronomy \& Astrophysics, P.O. Box 23-141, Taipei 10617, Taiwan}
\affil{$^{13}$Department of Physics, New Mexico Institute of Mining and Technology, 801 Leroy Place, Soccoro, NM 87801, USA}
\affil{$^{14}$Division of Physics, Faculty of Pure and Applied Sciences, Tsukuba, Ibaraki 305-8571, Japan}
\affil{$^{15}$Department of Physics, Ehime University, 2-5 Bunkyo-cho, Matsuyama, Ehime 790-8577, Japan}

\begin{abstract}
We present an unprecedented measurement of the disc stability and local instability scales in the luminous infrared Seyfert 1 host, NGC\,7469, based on ALMA observations of dense gas tracers and with a synthesized beam of $165\times 132$ pc. While we confirm that non-circular motions are not significant in redistributing the dense interstellar gas in this galaxy, we find compelling evidence that the dense gas is a suitable tracer for studying the origin of its intensely high-mass star forming ring-like structure. Our derived disc stability parameter $\mathcal{Q}$ accounts for a thick disc structure and its value falls below unity at the radii in which intense star formation is found. Furthermore, we derive the characteristic instability scale $\lambda_c$ and find a striking agreement between our measured scale of $\sim 180$ pc, and the typical sizes of individual complexes of young and massive star clusters seen in high-resolution images. 
\end{abstract}

\keywords{galaxies: ISM --- galaxies: kinematics and dynamics --- galaxies: individual (NGC 7469)}

\section{Introduction}
\label{intro}
The central kpc region of the barred Seyfert 1 host NGC\,7469 displays a number of morphological and kinematic features that provide interesting clues on the dynamical behaviour of the interstellar medium and its connection with star formation across the disc in this galaxy. Previous high resolution studies carried out by \cite{Meixneretal1990,Salamanca1995,Genzeletal1995,Daviesetal2004,Daviesetal2005} and \cite{Hicksetal2009} have shown that the gas rich central region displays intense star formation in a $R\sim 0.5-1$ kpc ring-like structure contributing to up to 2/3 of the galaxy's bolometric luminosity ($L_\mathrm{bol} = 3\times10^{11} L_\odot$). The ring-like structure is dominated by young ($<100$ Myr) and massive ($10^5$-$10^7\ M_\odot$) super star clusters, with the enclosed mass of $\sim 2.7 \times 10^9\ M_\odot$ \citep{Mauderetal1994,DiazSantos2007}. The individual clusters are grouped in $\sim$ 150 -- 200 pc-scale star forming complexes \citep{DiazSantos2007}. The relatively gas poor nuclear 30 -- 60 pc region hosts a $<100$ Myr nuclear star cluster which contributes to $10\%$ of the nuclear bolometric luminosity and 30\% of the nuclear $K$-band continuum light. Furthermore, the intensities of the cold and warm molecular gas are consistent with a CO-to-H$_2$ conversion factor similar to other star forming galaxies, i.e., \xco = 0.4 -- 0.8 the conversion factor in the Milky Way \citep{Daviesetal2004,Bolatto2013}.

The dynamics of the molecular gas around the nucleus is consistent with a thin disc model with $\sim 10^8$ \Msun\ dynamical mass and $\sim 3100\ M_\odot$ pc$^{-2}$ surface density \citep{Hicksetal2009}. The stellar structure inside the nuclear 30 pc is dispersion dominated whereas the $V/\sigma$ rapidly raises at this radius, staying consistently around unity out to $\sim 150$ pc. Furthermore, \cite{Hicksetal2008} applied dynamical models to measure the mass of the central supermassive black hole to $\sim 5\times10^7$ \Msun.

\begin{figure}s
\centering\includegraphics[width=0.45\textwidth]{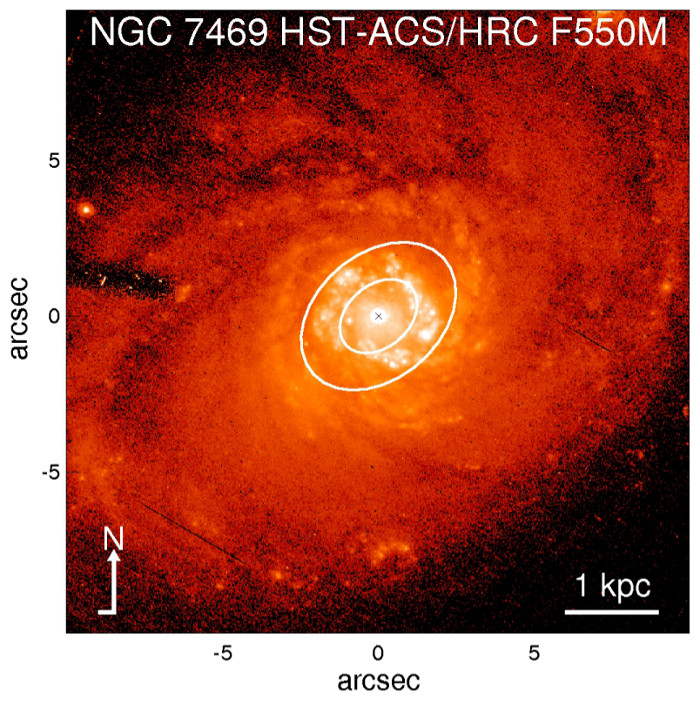}
\caption{The central region of NGC\,7469 with ellipses marking the 500 and 1000 pc boundaries in the plane of the galaxy disc.}
\label{fig:galaxy}
\end{figure}

The molecular bar-like structure observed in CO maps taken with the IRAM millimeter interferometer on the Plateau de Bure shown no clear signatures of bar kinematics \citep{Daviesetal2004} and it does not have a counterpart in high resolution $V$ or $K$-band images or radio continuum maps analyzed by \cite{Genzeletal1995,Mulchaeyetal1997,Laietal1999,Scovilleetal2000} and \cite{Colinaetal2001} (see also Fig.~\ref{fig:galaxy}). Associating the molecular bar-like morphology with a non-axisymmetric perturbation raises the problem of locating the position of the intensely star-forming ring-like structure. With its dominant recent burst of star formation, it would be likely placed in, or near, an associated inner resonance radius \citep[e.g.,][]{ReganTeuben2003,NPF2014,Lietal2015}.

Given that no significant kinematic characteristics of a barred potential has been found by \cite{Daviesetal2004}, it has been suggested that what appears to be a bar is likely a blending of gas clumps. Nevertheless, these authors assumed epicyclic approximation to associate the location of the ring-like structure with the inner Lindblad resonance radius of a primary bar, rotating at a pattern speed of $<80$ \kmskpc. Accordingly, the pattern speed for the region inside the ring was estimated to $235\pm5$ \kmskpc. Congruent with this picture, the star-forming regions in NGC\,7469 would be supported by its non-axisymmetric structure(s). However, strong bar/spiral-induced flows would be expected to sustain the large amounts of molecular gas (gas-to-dust ratio $\sim 600$) with \HCNJ/\COJ\ $\sim 1$ \citep{Papadopoulos2012}. 

While these results provide important clues on the fate of the interstellar gas in NGC\,7469, a concise picture for the dynamical effects that trigger the intense star formation in this galaxy is missing. To build a realistic scenario for the relative role of the local and global instabilities that ignite these intense bursts of star formation in the central kpc radius of NGC\,7469, we present here a detailed analysis of the distribution and kinematics of the densest phase of the molecular gas in the region of interest. We use the \HCOJ\ and \HCNJ\ emission lines at high spatial and velocity resolution to derive the instability parameter and the characteristic wavenumbers that describe the local instability scales.

Throughout this work, we assume NGC\,7469 to be at an inclination of $45\deg$ and a distance of 68 Mpc \citep{RC3}, implying 330 pc/\arcsec, and consistent with the above mentioned studies. The high angular resolution of Atacama Large Millimeter/sub-millimeter Array (ALMA) allows us to characterize the stability of the molecular gas throughout the central kpc region in this prototype luminous infrared galaxy. We show for the first time, that it is possible to directly quantify the stability criteria that lead to the star formation in the ring-like structure, accounting for up to 2/3 of the galaxy's bolometric luminosity.

\begin{figure*}
\centering\includegraphics[width=0.90\textwidth]{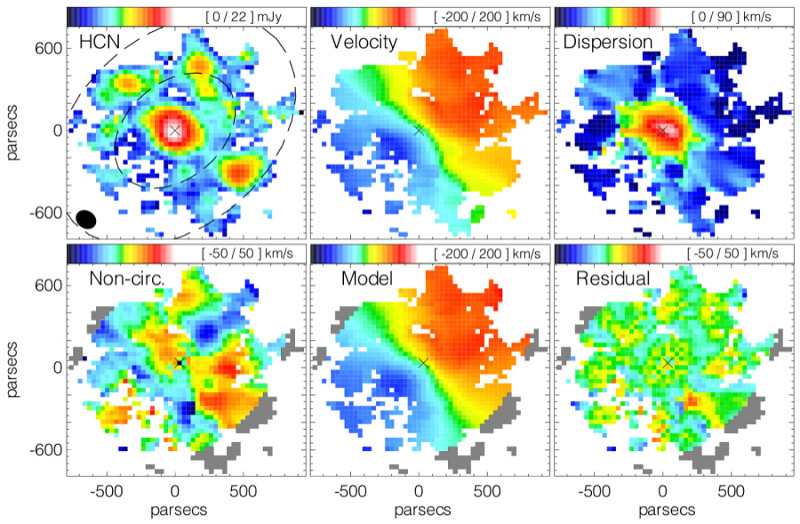}
\caption{\HCNJ\ distribution and kinematics (top), and a representation of the noncircular velocities with a harmonic model to quantify their amplitudes and a final residual (data minus harmonic model) at the bottom row. The maps in the top row are results of single Gaussian fits to the individual spectra. The bottom left panel shows the marginal amplitudes of the non-circular motions which are results from the fit to the observed velocity field. The cross marks the position of the dynamical centre. The filled ellipse in the top left panel shows the synthesized beam size, and the dashed ellipses mark the ring-like structure at $R\sim$ 500 -- 1000 pc. These maps remain consistent for the \HCOJ\ tracer, but slightly noisier.}
\label{fig:2Dmaps}
\end{figure*}

\section{Observations, Maps, and Profiles}
\label{sec:maprofiles}
We observed NGC\,7469 with ALMA on November 3rd and 4th, 2013, using 28 antennas in C32-2 configuration (Proposal ID 2012.1.00165.S, PI: Takuma Izumi). With 18\arcsec\ primary beam size and $0\farcs5\times 0\farcs4$ synthesized beam size, sampled at $0\farcs1$ pixels$^{-1}$, ALMA Band-7 receiver was tuned to cover the redshifted lines of \HCNJ\ ($\nu_\mathrm{rest}= 354.505$ GHz) and \HCOJ\ ($\nu_\mathrm{rest}= 356.734$ GHz). The original velocity resolution was  0.43 \kms\ but to achieve the amplitude-over-noise level $\geq 2$, we binned the data at an effective resolution of 20.2 \kms\ \citep[see][for further details]{Izumi1}.

To derive the gas distribution and kinematic information, we applied single Gaussian fits to each individual spectrum in the data cube, keeping only the positive-amplitude fits. Comparison between Gaussian fitting results and conventional moment maps confirm that the line fluxes, line-of-sight velocities and velocity dispersions ($\sigma$) as resulting from Gaussian fits are all reliable and fully consistent with the maps presented in \cite{Izumi1}. The difference is that our fitting scheme allows retrieving maps with a wider spatial coverage and recovering more extended emission, making our maps more suitable for the analysis of the gas kinematics. We further estimated the contribution of the side-lobes to $< 15\%$ at $R<3\arcsec$, hence not large enough to distort the derived kinematics (see Fig.~\ref{fig:2Dmaps}). While the emission line amplitudes slightly vary for the different tracers, we note that their morphology remains unchanged. 

Galactocentric flux and $\sigma$ profiles are derived applying robust statistics, ensuring reliable results with up to $50\%$ tolerance level. Moreover, the above estimated kinematics depend on the implicit assumption that the data comprise a random sample from a normal distribution. Noisy values or regions with stronger presence of a secondary component are thus accounted for by using robust statistics. Accordingly, we cut each two-dimensional map into projected circular galactocentric rings. At each radius (in the plane of the galaxy) we derive the median value for the flux and $\sigma$, followed by the error in the derived median values as described in \cite{HuberRonchetti2009}.

Numerous tests with varying ring widths and ring radii confirm that the exceptional quality and broad dynamical range of the ALMA data allows a stable median and error estimation even when median fluxes and $\sigma$ values are derived at radii a factor two smaller than the synthesized beam size. The smallest reliable step is thus $0\farcs25$, corresponding to an increment value of 82.5 pc, and this will be the step size adopted throughout the analysis presented here. Furthermore, the median velocity error across the field is 5.5 \kms, and the median emission line amplitude error across the field is 0.06 mJy beam$^{-1}$ \kms. All these parameters remain within the uncertainties both for the \HCN\ and the \HCO\ lines.

\begin{figure}
\centering\includegraphics[width=0.45\textwidth]{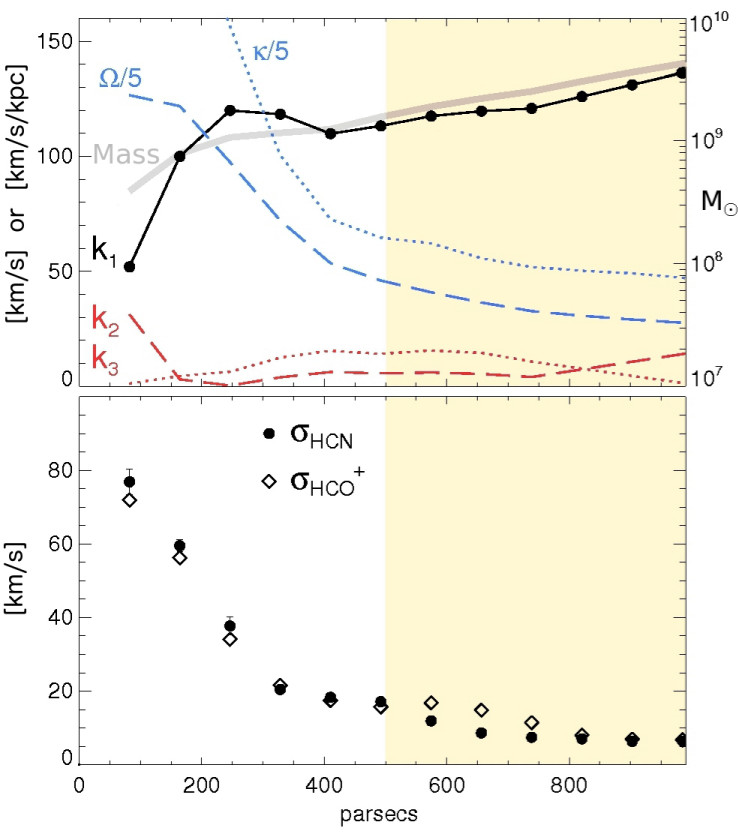}
\caption{The one-dimensional profiles for the best fit rotation curve $k_1$ and corresponding angular frequency $\Omega$ and epicyclic frequency $\kappa$, and the non-circular motions up to and including the third order (top panel), all in the plane of the sky. Both frequency curves are defined in section~\ref{sec:maprofiles} and they are here divided by five for illustration purposes. The dynamical mass (thick gray curve) accounts for the rotation curve, the \HCN\ $\sigma$ values shown in the bottom panel, and the assumed  $45\deg$ inclination. The \HCO\ $\sigma$ profile is shown in black diamonds. The shaded area marks the ring-like structure.}
\label{fig:vsigma}
\end{figure}

At these radial steps, we also calculate the rotation curve by assuming that the circular rotation is the dominant kinematic feature and that our measurements refer to positions on a single inclined disk. We then use the method described in \citep[e.g.,][]{Schoenmakers1997,Fathietal2005,vdVF2010} to derive the best representation of the rotation curve. Accordingly, the line-of-sight velocity is written as the harmonic series $k_0(R) + \sum_{m=1}^n k_m(R) \cos\left(m[\psi-\psi_m(R)]\right)$, where $k_0$ is the systemic velocity, $k_1$ is the best fit circular velocity, $R$ is the galactocentric radius, $k_m|_{m>1}$ are higher order velocity amplitudes and $\psi_m$ are phase shifts \citep[see also][ and Fig.~\ref{fig:vsigma} here]{vdVF2010}. Adopting $45\deg$ inclination, we find that the mean systemic velocity is $4860$ \kms\  and the kinematic position angle is $132\deg$. Once the rotation curve is derived at each radius $R$, we calculate the angular frequency $\Omega = k_1/R$ and the epicyclic frequency $\kappa = \sqrt{R d\Omega^2/dR + 4\Omega^2}$ curves (see Fig.~\ref{fig:vsigma}).

\section{Results and Discussion}
Rotation dominates the kinematics of the disc and our data show that both the \HCN\ and the \HCO\ lines trace the same dynamical entity. Henceforth, we use the \HCN, which provides less noisy maps, to derive the dynamical properties of the dense interstellar component, associated with the high-mass star forming gas in this galaxy.

\subsection{Dynamical Mass}
The low central $V/\sigma$ ($\sim 0.5$) found by \cite{Hicksetal2009} indicates that the random motions play a significant role in the central parts of NGC\,7469. Our Fig.~\ref{fig:vsigma} also confirms that the random motions have to be accounted for when deriving the dynamical mass. We assume that the \HCN\ $\sigma$ rises due to symmetric macroscopic motions and we calculate the dynamical mass in the central kpc radius applying the equation $M = R(k^{2}_{1}+3\sigma^2)/G$ \citep{Benderetal1992}, where $R$ is radius, $k_1$ is the best representation of the rotation curve (as described above) and $G$ is the gravitational constant (see Fig.~\ref{fig:vsigma}). The mass curve  shown in Fig.~\ref{fig:vsigma} is in full agreement with the measurements listed in \cite{Genzeletal1995} and \cite{Hicksetal2009}, and at the ring-like feature, our derived mass can encompass the combined masses for the star forming regions, as derived by \cite{DiazSantos2007}.

\begin{figure}
\centering\includegraphics[width=0.45\textwidth]{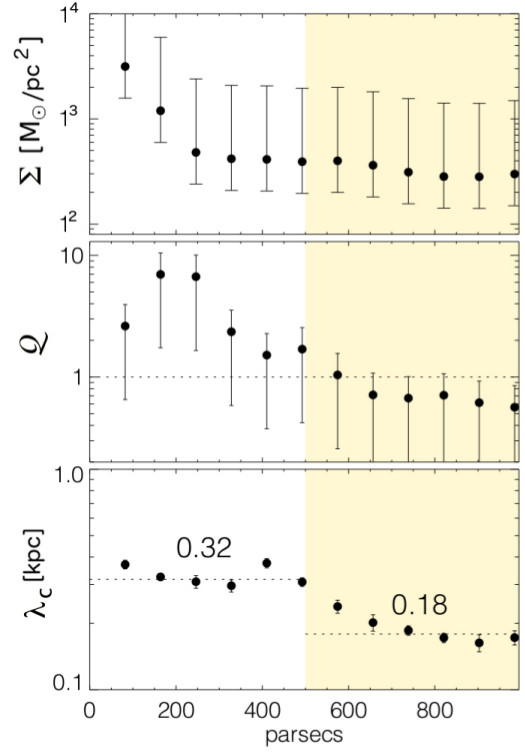}
\caption{\HCN\ surface density $\Sigma$, stability parameter $\mathcal{Q}$ and instability scale $\lambda_c$ (top, middle and bottom panel, respectively). The two dashed lines in the bottom panel mark the median value within each region, as indicated. The shaded area marks the ring-like structure of intense star formation. The uncertainties in $\Sigma$ (as described in section~\ref{sec:density}) are taken into account when deriving $\mathcal{Q}$ (see also section~\ref{sec:stability}).}
\label{fig:xhcn}
\end{figure}

\subsection{Surface Density of the Dense Gas Phase}
\label{sec:density} 
Considering the narrow range of the critical densities and the similarity of the line shapes, it is reasonable to conclude that the \HCN\ and \HCO\ are both emitted from the same volume \citep{Wadaetal2009,Zhangetal2014}, associated with the dense cores of star forming giant molecular clouds \citep{Kripsetal2008,Greveetal2009,AndrewsThompson2011}. Modeling by \cite{Meijerinketal2007} and \cite{Kazandjianetal2012} determined the thermal and chemical balance in similar environments and found that the densities in these regions vary between $10^4$ and $10^6$ cm$^{-3}$. In particular, where HCN/HCO$^+$ ratio is greater than unity, the most likely number density is $n_\mathrm{(H_2)}\sim 10^5$ cm$^{-3}$. NGC\,7469 fulfills this criterion \citep{Izumi1}. 

In Arp\,220-like environments, \cite{Greveetal2009} found a wide range of kinetic temperatures $T_k=10-120$ K. Moderately star forming galaxies exhibit values closer to the lower end of this temperature range \citep[e.g.,][]{GS2004,Kripsetal2008,Imanishietal2010}, hence, we find it reasonable to assume $T_k=25$ K in NGC\,7469.

Using the above density and temperature throughout the entire region studied here, we calculate the conversion factor \xhcn$=2.1 \times \sqrt{n_\mathrm{(H_2)}}/T_b \sim 27\ M_\odot$ (K \kms\ pc$^2$)$^{-1}$ \citep{Kripsetal2008}. Combined with the ratio \HCNRatio = $0.6$ \citep{Viti2014} we convert the \HCN\ flux into  molecular gas surface density $\Sigma$ (in $M_\odot$ pc$^{-2}$). We note that the uncertainties in the density, temperature and \HCNRatio\ are large enough to dominate the uncertainties in the $\Sigma$ values illustrated in Fig.~\ref{fig:xhcn}. Bothe $\Sigma$ and disc stability parameter (as described in section~\ref{sec:stability}) depend highly on the assumed temperature and number density. Decreasing the \HCNRatio\ will linearly increase the density values, while increasing the temperature will decrease the conversion factor. 

Given that the active galactic nucleus in NGC\,7469 only influences the inner $<40$ pcs \citep{Viti2014,Izumi2}, our $\Sigma$ curve is not affected by the corresponding radiation field. We further note that our measurements are prone to beam dilution up to a factor 5 \citep[e.g.,][]{Sakamotoetal2011} and that this effect would translate into a similar level of underestimation of the $\Sigma$. Hence, we provide a rough estimate of the uncertainties for the $\Sigma$ and subsequent stability parameter, using a factor 5 as upper value, and a factor two higher kinetic temperature \citep[e.g.,][]{ScovilleALMA2015} as lower $\Sigma$ limit (Fig.~\ref{fig:xhcn}).

\subsection{Disc Stability Analysis}
\label{sec:stability}
To analyze the disc stability in the central $R\sim 1$ kpc of NGC\,7469, we assume that the velocity dispersion is isotropic and we derive the Toomre-like stability parameter $\mathcal{Q}$ for the \HCN, taking into account also the disc thickness \citep{RomeoWiegert2011,RomeoFalstad2013}. Accordingly, $\mathcal{Q}=3\kappa\sigma/2\pi G\Sigma$, where all the parameters are described above. \cite{RomeoWiegert2011} have further shown that the dispersion relation for the stability parameter $\mathcal{Q}$ converges to Jeans dispersion relation. The middle panel of Fig.~\ref{fig:xhcn} illustrates that while the stability parameter $\mathcal{Q}$ is consistently above unity where no star formation is observed, this parameter decreases below unity at the region of the ring-like structure in  NGC\,7469. The corresponding error bars account for the uncertainties in the surface density $\Sigma$ and the errors in the derived median $\sigma$ values. 

This is a strong result, since it shows that even the densest gas, which is likely sunk to the disk mid-plane \citep{MO2007} and least prone to perturbations, is unstable in the ring-like structure. The instability gives rise to the intense formation of the high-mass young stars that have been reported in the literature \citep{DiazSantos2007}. Another interesting point here is that the \HCN\ $\mathcal{Q}$ highlights the role of local gravitational instabilities which lead to the intense star formation in this galaxy. \cite{RomeoFathi2015} presented a detailed discussion on the physical phenomena that could contribute to the high $\mathcal{Q}$ at small radii. Moreover, while gas turbulence alters the condition for star-gas decoupling and increases the least stable wavelength, it hardly modifies the Q parameter at scales larger than about 100 pc \citep{HoffmannRomeo2012,Agertzetal2015}. Given the metallicity of the central kpc radius of NGC\,7469, it is possible that CO-dark molecular gas, anisotropy of the gas velocity dispersion and non-axisymmetric perturbations all contribute to the disc stability at $R \lesssim 500$ pc region, by up to 50\% or so. However, outside this radius, the low velocity dispersion of the \HCN\ suggests that the contribution of the CO-dark gas and anisotropy may be marginal \citep{RomeoFathi2015}. Furthermore, \cite{Sanietal2012}  found that the Toomre parameter traced by \HCN\ is above unity at the centre of other Seyfert galaxies.

The prescription of \cite{RomeoFalstad2013} further allows measuring the characteristic instability scale $\lambda_c$, i.e. the physical scales at which a realistically thick galactic disc becomes locally unstable as $\mathcal{Q}$ drops below unity. It is defined as $\lambda_c = 2\pi \sigma/\kappa$ and deriving the scales for local perturbations does not exclude the role of global perturbation in reshaping the disc in NGC\,7469. Indeed, a locally stable disc can still be globally unstable to bi-symmetric gravitational perturbations. Moreover, the dynamics and evolution of disc structures depend critically on the radial profile of the $\mathcal{Q}$ stability parameter \citep[][and references therein]{RomeoFalstad2013}. 

The lower panel in Fig.~\ref{fig:xhcn} shows a marked difference between $\lambda_c$ interior to and on the ring-like structure in NGC\,7469. It is striking that in this region our measured $\sim 180$ pc scale (median $\lambda_c = 0.18$ kpc) is fully comparable to the measured sizes for the star forming complexes. The derived local instability scale for the densest gas is further supported by the observed young (5 -- 6 Myr) and highly obscured ($A_V\sim 10$ mag) population of stars that coexist with an intermediate age (15 -- 35 Myr) less extincted population \citep{DiazSantos2007}. The young and massive population accounts for $\sim 1/3$ of the total stellar mass and up to 2/3 of the infrared luminosity in the ring-like structure \citep{DiazSantos2007}.

\section{Conclusions}
Studies of star formation in discs suggest that massive stars are predominantly formed in the densest cores of giant molecular clouds \citep[e.g.,][]{Krumholz2014}, hence it is commonly assumed that \HCN\ and \HCO\ trace these dense cores \citep{Wuetal2005,Ladaetal2012}. Due to the high critical densities, a tighter correlation is found between the intensities of these molecules and the far infrared emission \citep{GS2004}. This correlation further extends to dense cores undergoing high mass star-formation \citep{Ladaetal2012,Zhangetal2014}. 

The galactic scale intense star formation in NGC\,7469 is likely due to the combination of a past interaction with the neighboring IC\,5283 and the presence of its bar or spiral arms. However, the expected non-circular motions are not detected in our ALMA kinematic maps for this galaxy \citep[c.f.][]{Fathietal2005,Fathietal2006,Fathietal2013}. Our ALMA observations of \HCNJ\ and \HCOJ\ lines reveal a number of interesting clues on how local gravitational instabilities help forming stars in this galaxy.

We derive the radial profiles for the disc stability parameter $\mathcal{Q}$ and the characteristic instability scale $\lambda_c$ in NGC\,7469. We find signatures of disc instability on the ring-like structure, inside which, the disc is stable against star formation. At the highly star forming ring-like structure, the median local instability scale is $\sim 180$ pc, in full agreement with the sizes of individual star forming complexes seen in Hubble Space Telescope images. Our stability analysis accounts for a disc structure with finite thickness, and as discussed in \cite{RomeoFathi2015} and \cite{Agertzetal2015}, turbulence could alter the condition for star-gas decoupling and increases the least stable wavelength, but hardly modifies the stability parameter $\mathcal{Q}$ at scales larger than about 100 pc.

Our robust results further show that short-wavelength approximation is satisfied in the region studied here and that the ring-like structure could in essence be only a tightening of the galactic-scale non-axisymmetric spiral arms. It will be interesting to apply this analysis to the warmer, less dense and more diffuse CO gas, however, as it was pointed out by \cite{RomeoFalstad2013} such a comparison must be based on identical data quality and resolution. At the moment we have no CO data at hand that match the resolution, sensitivity, and spatial coverage of the ALMA observations analyzed here.

\section*{ACKNOWLEDGMENTS}
We thank the anonymous referee for insightful comments which helped improving our manuscript. This paper makes use of the following ALMA data: ADS/JAO.ALMA\#2012.1.00165.S. ALMA is a partnership of ESO (representing its member states), NSF (USA) and NINS (Japan), together with NRC (Canada), NSC and ASIAA (Taiwan), and KASI (Republic of Korea), in cooperation with the Republic of Chile. The Joint ALMA Observatory is operated by ESO, AUI/NRAO and NAOJ. SM is supported by MoST 103-2112-M-001-032-MY3. KF acknowledges the hospitality of ESO in Garching, where part of this work was carried out.

\label{lastpage}

\begin{thebibliography}{} 

\bibitem[Agertz et al. (2015)]{Agertzetal2015}
Agertz, O., Romeo, A. B., Grisdale, K. 2015, MNRAS, 449, 2156

\bibitem[Andrews \& Thompson (2011)]{AndrewsThompson2011}
Andrews, B. H., Thompson, T. A. 2011, ApJ, 727, 97

\bibitem[Bender et al. (1992)]{Benderetal1992}
Bender, R., Burstein, D., Faber, S. M. 1992, ApJ, 399, 462

\bibitem[Bolatto et al. (2013)]{Bolatto2013}
Bolatto, A. D., Wolfire, M., Leroy, A. K. 2013, ARA\&A, 51, 207

\bibitem[Colina et al. (2001)]{Colinaetal2001}
Colina, L. et al. 2001, ApJ, 553, L19

\bibitem[Davies et al. (2004)]{Daviesetal2004}
Davies, R., Tacconi, L., Genzel, R. 2004, ApJ, 602, 148

\bibitem[Davies et al. (2005)]{Daviesetal2005}
Davies, R. et al. 2005, ApJ, 633, 105

\bibitem[de Vaucouleurs et al. (1991)]{RC3}
de Vaucouleurs, G., et al. 1991, Third Reference Catalogue of Bright Galaxies, Springer, New York, NY (USA)

\bibitem[D\'{\i}az-Santos et al. (2007)]{DiazSantos2007}
D\'{\i}az-Santos, T. et al. 2007, ApJ, 661, 149

\bibitem[Fathi et al. (2005)]{Fathietal2005}
Fathi, K. et al. 2005, MNRAS, 364, 773

\bibitem[Fathi et al. (2006)]{Fathietal2006}
Fathi, K. et al. 2006, ApJ, 641, L25

\bibitem[Fathi et al. (2013)]{Fathietal2013}
Fathi, K. et al. 2013, ApJ, 770, L27

\bibitem[Gao \& Solomon (2004)]{GS2004}
Gao, Y., Solomon, P. M. 2004, ApJ, 606, 271

\bibitem[Genzel et al. (1995)]{Genzeletal1995}
Genzel, R. et al. 1995, ApJ, 444, 129

\bibitem[Greve et al. (2009)]{Greveetal2009}
Greve, T. R. et al. 2009, ApJ, 692, 1432

\bibitem[Hicks \& Malkan (2008)]{Hicksetal2008}
Hicks, E. K. S., \& Malkan, M. A. 2008, ApJS, 174, 31

\bibitem[Hicks et al. (2009)]{Hicksetal2009}
Hicks, E. K. S. et al. 2009, ApJ, 696, 448

\bibitem[Hoffmann \& Romeo (2012)]{HoffmannRomeo2012}
Hoffmann, V., Romeo, A. B. 2012, MNRAS, 425, 1511

\bibitem[Huber \& Ronchetti (2009)]{HuberRonchetti2009}
Huber, P. J., Ronchetti, E. M. 2009, "Robust Statistics", Wiley Series in Probability and Statistics

\bibitem[Imanishi et al. (2010)]{Imanishietal2010}
Imanishi, M. et al. 2010, PASJ, 62, 201

\bibitem[Izumi et al. (2015a)]{Izumi1}
Izumi, T. et al. 2015a, ApJ, submitted

\bibitem[Izumi et al. (2015b)]{Izumi2}
Izumi, T. et al. 2015b, ApJ, submitted

\bibitem[Kazandjian et al. (2012)]{Kazandjianetal2012}
Kazandjian, M. V. et al. 2012, A\&A 542, 65

\bibitem[Krips et al. (2008)]{Kripsetal2008}
Krips, M. et al. 2008, ApJ, 677, 262

\bibitem[Krumholz (2014)]{Krumholz2014}
Krumholz, M. R. 2014, in ASSL, 412, 43

\bibitem[Lada et al. (2012)]{Ladaetal2012}
Lada, C. J. et al. 2012, ApJ, 669, 289

\bibitem[Lai et al. (1999)]{Laietal1999}
Lai, O., Rouan, D., Alloin, D. 1999, in Astronomy with Adaptive Optics, ed. Bonaccini (Garching: ESO), 555

\bibitem[Li et al. (2015)]{Lietal2015}
Li, Z., Shen, J., Woong-Tae, K. 2015, ApJ, submitted (preprint available at arXiv:1503.02594)

\bibitem[Mauder et al. (1994)]{Mauderetal1994}
Mauder, W. et al. 1994, A\&A, 285, 44

\bibitem[McKee \& Ostriker (2007)]{MO2007}
McKee, C. F., Ostriker, E. C. 2007, ARA\&A, 45, 565

\bibitem[Meijerink et al. (2007)]{Meijerinketal2007}
Meijerink, R., Spaans, M., Israel, F. P. 2007, A\&A, 461, 793

\bibitem[Meixner et al. (1990)]{Meixneretal1990}
Meixner, M. et al. 1990, ApJ, 354, 158

\bibitem[Mulchaey et al. (1997)]{Mulchaeyetal1997}
Mulchaey, J. S., Regan, M. W., Kundu, A. 1997, ApJS, 110, 299

\bibitem[Papadopoulos et al (2012)]{Papadopoulos2012}
Papadopoulos, P. P. et al. 2012, MNRAS, 426, 2601

\bibitem[Pi\~{n}ol-Ferrer et al. (2014)]{NPF2014}
Pi\~{n}ol-Ferrer, N. et al. 2014, MNRAS, 438, 971

\bibitem[Regan \& Teuben (2003)]{ReganTeuben2003}
Regan, M. W., Teuben, P. 2003, ApJ, 582, 723

\bibitem[Romeo \& Wiegert (2011)]{RomeoWiegert2011}
Romeo, A. B., Wiegert, J. 2011, MNRAS, 416, 1191

\bibitem[Romeo \& Falstad (2013)]{RomeoFalstad2013}
Romeo, A. B., Falstad, N. 2013, MNRAS, 433, 1389

\bibitem[Romeo \& Fathi (2015)]{RomeoFathi2015}
Romeo, A. B., Fathi, K. 2015, MNRAS, in press (preprint available at arXiv:1503.01326)

\bibitem[Sakamoto et al. (2011)]{Sakamotoetal2011}
Sakamoto, K. et al. 2011, ApJ, 735, 19

\bibitem[Salamanca et al. (1995)]{Salamanca1995}
Salamanca, I. et al. 1995, A\&ASS, 111, 283

\bibitem[Sani et al. (2012)]{Sanietal2012}
Sani, E. et al. 2012, MNRAS, 424, 1963

\bibitem[Schoenmakers et al. (1997)]{Schoenmakers1997}
Schoenmakers, R. H., Franx, M., de Zeeuw, P. T. 1997, MNRAS, 292, 349

\bibitem[Scoville et al. (2000)]{Scovilleetal2000}
Scoville, N., et al. 2000, AJ, 119, 991

\bibitem[Scoville et al. (2015)]{ScovilleALMA2015}
Scoville, N. et al. 2015, ApJ, 800, 70

\bibitem[van de Ven \& Fathi (2010)]{vdVF2010}
van de Ven, G., Fathi, K. 2010, ApJ, 723, 767

\bibitem[Viti et al. (2014)]{Viti2014}
Viti, S. et al. 2014, A\&A, 570, 28

\bibitem[Wada et al. (2009)]{Wadaetal2009}
Wada, K., Papadopoulos, P. P., Spaans, M. 2009, ApJ, 702, 63

\bibitem[Wu et al. (2005)]{Wuetal2005}
Wu, J. et al. 2005, ApJ, 635, L173

\bibitem[Zhang et al. (2014)]{Zhangetal2014}
Zhang, Z. et al. 2014, ApJ, 784, 31

\end{thebibliography}
\end{document}